%% file: paper.tex
\newif\ifsandy
\sandyfalse

\ifsandy
    \documentclass[letterpaper,10pt]{article}
\else
    \documentclass[letterpaper,twocolumn,10pt]{article}
    \usepackage{usenix2019_v3}
\fi

\usepackage{epsfig,endnotes}

\usepackage{fancyhdr}
\usepackage{amsmath}
\usepackage{booktabs}
\usepackage{graphicx}
\usepackage{setspace}
\usepackage{xspace}
\usepackage{caption}
\usepackage{subcaption}
\usepackage{textcase}
\usepackage{textcomp}
\usepackage{multirow}
\usepackage{makecell}

\usepackage{dingbat}
\newcommand{\addfigure}[1]{\centering\includegraphics[width=2.8in]{figures/#1}\vspace{-.1in}}

\everypar=\expandafter{\the\everypar\looseness=-1 }
\usepackage[draft]{todonotes}   
\usepackage{enumitem}
\setlist{nolistsep}

\newcommand{\squishlist}{
   \begin{list}{$\bullet$}
    { \setlength{\itemsep}{4pt}      \setlength{\parsep}{3pt}
      \setlength{\topsep}{3pt}       \setlength{\partopsep}{0pt}
      \setlength{\leftmargin}{1.5em} \setlength{\labelwidth}{1em}
      \setlength{\labelsep}{0.5em} } }

\newcommand{\squishlisttwo}{
   \begin{list}{$\bullet$}
    { \setlength{\itemsep}{0pt}    \setlength{\parsep}{0pt}
      \setlength{\topsep}{0pt}     \setlength{\partopsep}{0pt}
      \setlength{\leftmargin}{2em} \setlength{\labelwidth}{1.5em}
      \setlength{\labelsep}{0.5em} } }

\newcommand{\squishend}{
    \end{list}  }

\newif\ifmarnote
\marnotefalse 
\newcommand{\marnote}[1]{%
\ifmarnote%
\setlength{\marginparwidth}{\oddsidemargin}%
\addtolength{\marginparwidth}{.8in}
\addtolength{\marginparwidth}{-\marginparsep}%
\marginpar{\color{red}\raggedright \scriptsize \em #1}%
\fi}

\newcommand{\agate}{Agate\xspace}
\newcommand{\safe}{\textsc{safe}\xspace}
\newcommand*\dash{\unskip\kern.16667em---\penalty\exhyphenpenalty
        \hskip.16667em\relax
}
\newif\ifanon
\anonfalse

\ifanon
  
\begin{document}
\else
  \pagestyle{fancy}
  \begin{document}
  \renewcommand{\headrulewidth}{0pt}
  \chead{}
  \lhead{}
  \rhead{}
\fi

\def\compactify{\itemsep=2pt \topsep=2pt \partopsep=1pt \parsep=1pt \leftmargin=1.2em}
\let\latexusecounter=\usecounter
\newenvironment{CompactEnumerate}
  {\def\usecounter{\compactify\latexusecounter}
   \begin{enumerate}}
   {\end{enumerate}\let\usecounter=\latexusecounter}
\newenvironment{CompactItemize}
   {\def\usecounter{\compactify\latexusecounter}
   \begin{itemize}}
   {\end{itemize}\let\usecounter=\latexusecounter}

\makeatletter
\def\tup{%
  \def\tupInside\tupStart##1,##2\tupEnd{%
    ##1
    \def\tempa{}\def\tempb{##2}
    \ifx\tempa\tempb\else
    ,\;\tupInside\tupStart##2\tupEnd
    \fi}
  \def\tupBraces##1{\left\langle\tupInside\tupStart##1,\tupEnd\right\rangle}
  \def\tupBrackets<##1>{\tupBraces{##1}}
  \@ifnextchar<\tupBrackets\tupBraces}
\def\set#1{%
  \def\setInside\setStart##1|##2\setEnd{%
    \def\tempa{}\def\tempb{##2}
    \ifx\tempa\tempb
    ##1
    \else
    \def\stripPipe####1|{####1}
    ##1 \mid \stripPipe##2
    \fi}
  \left\{\setInside\setStart#1|\setEnd\right\}}
\makeatother

\newcommand{\id}[1]{\textsc{\MakeTextLowercase{#1}}}
\newcommand{\attrib}[1]{\textsc{\MakeTextLowercase{#1}}}
\newcommand{\resource}[1]{\textsc{\MakeTextLowercase{#1}}}
\newcommand{\app}[1]{\textsc{\MakeTextLowercase{#1}}}
\newcommand{\principal}[1]{\textsc{\MakeTextLowercase{#1}}}
\newcommand{\exception}[2]{\resource{#1}.\textsc{\MakeTextLowercase{#2}}}
\newcommand{\group}[2]{\principal{#1}.\textsc{\MakeTextLowercase{#2}}}
\newcommand{\code}[1]{\texttt{\small #1}}
\newcommand{\ppp}[0]{P$^\text{3}$\xspace}

\newcommand{\captionfonts}{ }

\makeatletter  
\long\def\@makecaption#1#2{%
  \vskip 0.1in
  \sbox\@tempboxa{{\captionfonts #1: #2}}%
  \ifdim \wd\@tempboxa >\hsize
    {\captionfonts #1: #2\par}
  \else
    \hbox to\hsize{\hfil\box\@tempboxa\hfil}%
  \fi
  \vskip 0in}
\makeatother   








\title{ \Large \bf Making Distributed Mobile Applications \safe:\\
  Enforcing User Privacy Policies on Untrusted Applications\\with Secure Application Flow
  Enforcement}

\renewcommand\theadfont{\small\bf}
\author{
\ifanon
{
}
\else
{\rm Adriana Szekeres} \and {\rm Irene Zhang} \and {\rm Katelin
  Bailey} \and {\rm Isaac Ackerman} \and
  {\rm Haichen Shen} \and {\rm Franziska Roesner} \and 
  {\rm Dan R. K. Ports} \and {\rm Arvind Krishnamurthy} \and
  {\rm Henry M. Levy} \vspace{3pt} \and \newline
  {\rm University of Washington} \vspace{10pt} \\
\fi
}
\date{}
\maketitle

\ifsandy
    \doublespacing
\fi
\vspace{-1in}
\begin{abstract}
\input{abstract.tex}
\end{abstract}




\section{Introduction}
\label{sec:intro}
\input{intro}


\section{The \safe Framework}
\label{sec:overview}
\input{overview}


\section{Agate, a \safe Mobile OS}
\label{sec:os}
\input{os}

\section{Magma, a \safe Runtime System}
\label{sec:platform}
\input{platform}

\section{Geode, a \safe Storage Proxy}
\label{sec:storage}
\input{storage}

\section{Evaluation}
\label{sec:eval}
\input{eval}

\section{Related Work}
\label{sec:relwork}
\input{relwork}

\section{Conclusion}
\label{sec:conclusion}
\input{conclusion}

{\footnotesize \bibliographystyle{acm}




\bibliography{paper}}

\end{document}

%% file: abstract.tex

Today's mobile devices sense, collect, and store huge amounts of
personal information, which users share with family and friends
through a wide range of applications.  Once users give applications
access to their data, they must implicitly trust that the apps
correctly maintain data privacy.  As we know from both experience and
all-too-frequent press articles, that trust is often misplaced.

While users do not trust applications, they do trust their mobile
devices and operating systems.  Unfortunately, sharing applications
are not limited to mobile clients but must also run on cloud services
to share data between users.  In this paper, we leverage the trust
that users have in their mobile OSes to vet cloud services.

To do so, we define a new \emph{Secure Application Flow Enforcement}
(\safe) framework, which requires cloud services to attest to a system
stack that will enforce policies provided by the mobile OS for user
data.  We implement a mobile OS that enforces \safe policies on
unmodified mobile apps and two systems for enforcing policies on
untrusted cloud services.  Using these prototypes, we demonstrate that
it is possible to enforce existing user privacy policies on unmodified
applications.

%% file: intro.tex

Sharing is the hallmark of applications in the mobile era. Mobile
devices constantly collect information about their users (e.g., their
location, photos, etc.) and supply it to applications, which then
share this personal data with other users distributed over many mobile
devices. This data ranges from the mundane to the highly sensitive,
making protecting its privacy a critical challenge for modern
applications.

Mobile operating systems let users restrict application access to the
sensitive data on their devices (e.g., through the Android
Manifest~\cite{android_app_manif_devel_guide} or iOS privacy
settings~\cite{apple:_human_inter_guidel}).  However, once an app has
access, users must trust the app to ensure their privacy.  Almost all
apps offer their users a choice of privacy policies; unfortunately,
they frequently violate these policies due to
bugs~\cite{soze11:_mark_zucker_privat_photos_leaked, blue13:_resear,
  linked_hack} or other
reasons~\cite{notopoulos12:_snapc_featur_that_will_ruin_your_life,
  commission13:_android_flash_app_devel_settl,
  henry12:_twitt_is_track_you_web}.

While mobile OSes are effective at enforcing user privacy policies,
that enforcement does not extend to application backends and other
cloud services that sharing applications rely on to move data between
device.  Researchers have proposed distributed cloud platforms, but
they only support some application features~\cite{sangmin:pibox,
  popa14:mylar}, require application modification~\cite{giffin:hails,
  myers:jif, cheng12:aeolus} or have complex user
policies~\cite{zeldovich:dstar, stefan14:cowl}.
As a result, users are left to blindly trust that applications will
respect their privacy even as the apps move their data across a
complex landscape of backend servers, storage systems and cloud
services.

This paper offers a practical alternative for users.  Unlike existing
systems, we aim to enforce \emph{existing privacy policies on
  unmodified and untrusted sharing applications} across mobile devices
and cloud services.  We achieve this goal with a key insight: while
mobile OSes cannot enforce policies on cloud services, the OS can vet
cloud services on behalf of users and ensure that a cloud service will
respect a user's policies before handing over a user's data.  

We introduce a new \emph{Secure Application Flow Enforcement} (\safe)
framework for vetting systems that handle user data.  The framework
defines a single guarantee: \emph{Given a piece of user data and a
  \safe flow policy for that data, the system must ensure that it will
  only release that data and any data derived from that data to: (1)
  another \safe system or (2) an un-\safe system allowed by the \safe
  flow policy.}  \safe flow policies, detailed in
Section~\ref{sec:overview}, are access-control lists (ACLs) consisting
of users and groups that reflect existing policies already set by
users.

The \safe guarantee can be applied to any software that handles user
data, including systems, cloud services and user-facing applications.
However, it is clearly not practical to modify all applications to
meet the guarantee.  Instead, we rely on \emph{\safe enforcement
  systems}, trusted systems that ensure untrusted and unmodified
applications meet the \safe guarantee.  As a consequence, apps running
atop a \safe enforcement system can be deemed \emph{\safe apps}.

With this framework, users can begin by running a \emph{\safe
  enforcement operating system} on their mobile devices. Then, they
can trust the \safe OS -- and any untrusted apps running on it -- to
give sensitive user data to either another \safe system and
application or another user allowed within the \safe policy.  \safe
OSes vet untrusted application backends and cloud services by using
TPM-based attestation to verify that the cloud server is running a
trusted systems stack, including a trusted \safe enforcement system.
Once verified, the user's mobile OS can safely send sensitive user
data and be certain that the \safe guarantee will be upheld by the
\safe enforcement system and untrusted applications running on top.

The \safe guarantee is incredibly powerful: if a user gives a piece of
data to a \safe enforcement system along with a \safe policy, the user
can trust that the policy will be enforced no matter where the data
flows until it is released to another user allowed by the policy.  In
other words, the \safe framework lets users construct a chain of trust
from their mobile OS to an arbitrary set of cloud services to other
users' devices.  Furthermore, users do not need to understand which
cloud services their applications use, only to trust that the services
are verified \safe. In fact, users do not even need to understand the
\safe concept, provided that they trust their mobile OSes to be \safe
and to correctly verify that other systems that handle their data are
\safe as well.

The remainder of this paper describes the \safe framework
(\textsection\ref{sec:overview}) and the design and implementation of
three \safe enforcement systems:
\begin{itemize}
\item \emph{Agate}, a \safe mobile OS that securely collects user
  policies, enforces them on untrusted mobile apps and translates them
  to \safe policies for \safe cloud services
  (\textsection\ref{sec:os}).
\item \emph{Magma}, a \safe distributed cloud runtime system that
  enforces \safe policies on untrusted cloud backends using
  fine-grained information flow control
  (\textsection~\ref{sec:platform}).
\item \emph{Geode}, a \safe proxy that enforces \safe policies on
  untrusted storage systems which do not manipulate user data.
  (\textsection~\ref{sec:storage}).
\end{itemize}
Using these three systems, we demonstrate that it is possible to
enforce user policies end-to-end on unmodified distributed mobile apps
largely without changing the user experience.  We do so for several
existing applications, including a 70,000-line calendar application
and a 250,000-line chat application.  Furthermore, we show that the
\safe policy model supports a range of existing user policies and that
systems can implement \safe policy enforcement with little
user-noticeable overhead ($\approx$20\% on a mobile device).


%% file: overview.tex

This section summarizes the \safe framework, including the system
model and concepts, the policy model and the threat model.  We had
three goals when designing the framework: (1) minimize changes to user
experience, (2) minimize changes to application code, and (3) minimize
performance cost.  This section reviews the ways in which the \safe
design meets these goals.

\subsection{\safe System Model and Concepts}
\begin{figure*}[t]
  \centering
  \vspace{-.1in}
  \begin{minipage}{0.48\linewidth}
    \includegraphics[width=1.0\linewidth]{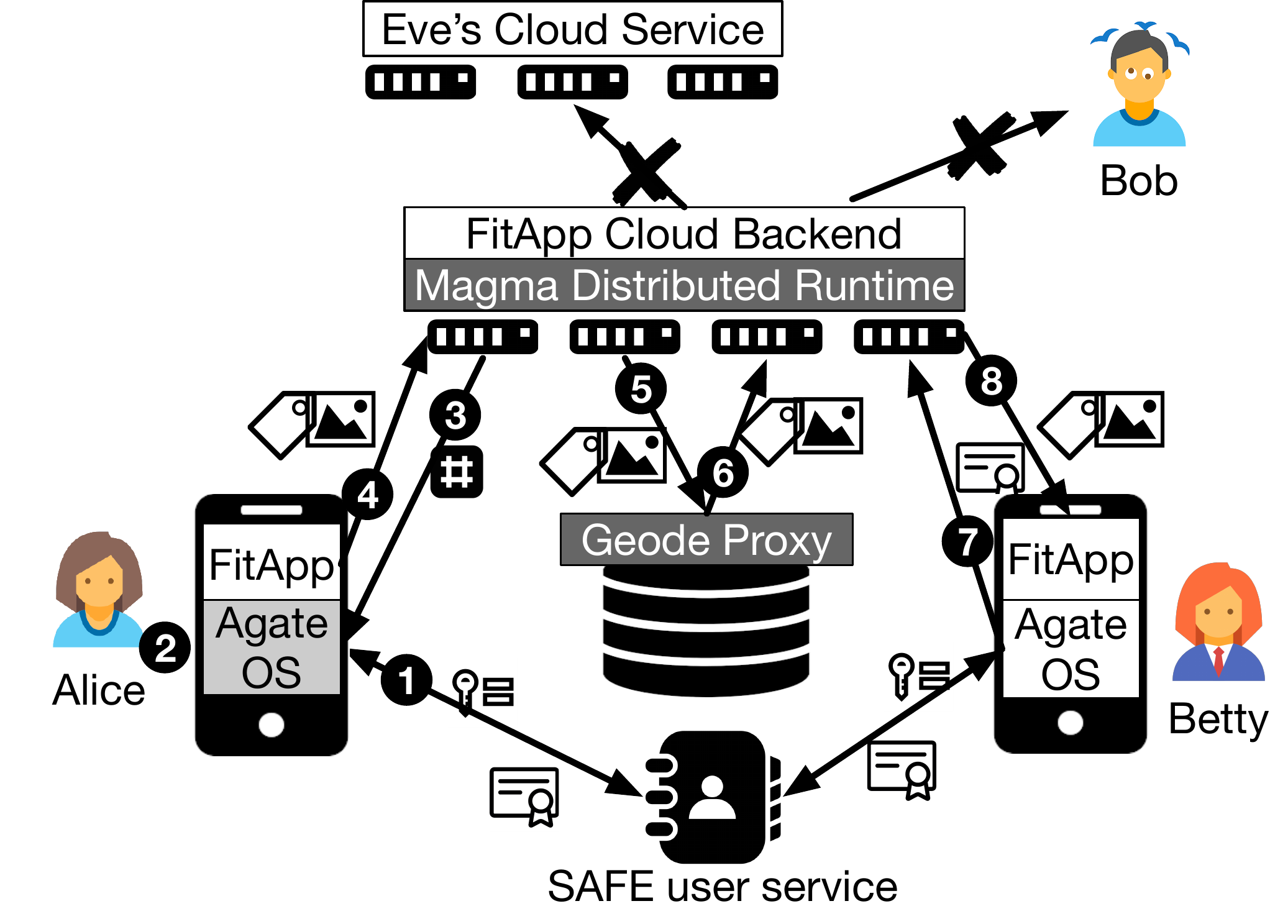}
  \end{minipage}
  \hspace{.2in}
  \begin{minipage}{0.48\linewidth}
    \begin{enumerate}
    \item Alice logs in to Agate to access FitApp
    \item Alice sets a policy in Agate to share her photos with Betty
      through FitApp
    \item Agate verifies that the FitApp cloud backend is a \safe
      service (it attests to running Magma)
    \item Agate allows the FitApp mobile client to send Alice's photo
      to the FitApp cloud backend
    \item Magma allows FitApp to store Alice's photo because the
      storage system attests to running the Geode Proxy
    \item Geode releases Alice's photo back to FitApp backend
    \item FitApp wants to send Alice's photo to Betty's phone, so
      Magma requests app and user certificates from Agate and checks
      the photo's \safe policy
    \item Magma allows FitApp to send Alice's photo to Betty's phone
    \end{enumerate}
  \end{minipage}
  \caption{\textbf{Example \safe ecosystem.}  We show the steps for
    Alice to share her photos with Betty securely through the \safe
    framework.  We color the systems based on Alice's trust in them;
    she trusts her \safe Agate OS running on her phone (light gray),
    she trusts her Agate to verify (using attestation) that the (dark
    gray) cloud systems are \safe, and she does not trust the white
    apps or systems.  Magma will not let FitApp send Alice's photo to
    Eve's cloud service because it is not an attested \safe cloud
    service.  Likewise, Magma ensures that FitApp cannot share Alice's
    photo with Bob because Bob is not in Alice's \safe policy. }
  \label{fig:arch}
\end{figure*}

Figure~\ref{fig:arch} shows the \safe system model.  A \emph{\safe
  application} consists of processes distributed across mobile devices
(i.e., mobile app clients) and cloud servers (i.e., cloud app
backends) as well as cloud services used by the application (e.g.,
distributed storage~\cite{13:_redis,fitzpatrick04:_distr}).  Some of
the data that a \safe application handles will have attached \safe
policies, while other data does not.  We assume mobile devices are
owned by users, while cloud servers are operated either by the
application provider (e.g., Twitter, Facebook) or by a third-party
cloud provider (e.g., Amazon).  The application runs on one or more
mobile OS platforms and one or more cloud platforms.

\subsubsection{\safe User Service}
Each \safe deployment instance, called an \emph{ecosystem}, is defined
by a centralized, trusted user management service.  The \emph{\safe
  user service} is shown in the bottom of Figure~\ref{fig:arch}.  The
\safe user service has two roles: (1) securely managing the principals
used to express \safe policies, including group membership management,
and (2) verifying app and user identities to authorize the release of
data to an untrusted device. The \safe user service gives every
application, user and group a unique identifier and stores a mapping
between names and ids.  It also authenticates users, issues
certificates to untrusted devices to authorize them to access data on
behalf of users and manages group membership.

Any trusted entity can launch a \safe user service and ecosystem to
support one or more applications.  Apps within an ecosystem can easily
exchange data and policies (if allowed by the \safe policy) because
they share the same \safe user service and its principals.  Apps
outside an ecosystem must negotiate a way to translate \safe policies
when exchanging data.

The \safe user service is not {\safe}-specific; there are many
existing single-sign-on (SSO) services that could be used to implement
the same functionality (e.g., Google Accounts~\cite{googl_accoun},
OpenID~\cite{openid}).  The only requirement is that the service be
able to authenticate users, issue certificates and manage group
membership.  For trust reasons, we assume that the \safe user service is
implemented and deployed by an entity separate from the application
provider; otherwise, we would have to trust the application to manage
users, which many fail at~\cite{linked_hack,17:_mohit_kumar}.  As an
example, Google could create its own ecosystem for Google apps by
deploying a \safe user service or using Google Accounts.

\subsubsection{\safe Mobile Operating Systems}
\marnote{say something about app certificates} Users run a \emph{\safe
  enforcement OS} on their mobile devices to ensure that user policies
are securely captured, enforced on untrusted mobile apps and expressed
to \safe cloud services.  The \safe OS also verifies cloud services as
being \safe before allowing apps to send sensitive user data.  This
paper describes the design of the Agate \safe OS (shown on Alice and
Betty's phones in Figure~\ref{fig:arch}), but we imagine that other
\safe OSes would exist.

Before running \safe apps, users must login to the \safe mobile OS,
which authenticates the user with the \safe user service.  The \safe
user service issues a user certificate to the \safe OS, which
authorizes it to collect data and policies on behalf of the logged-in
user and hold data shared with that user.  We assume that users trust
any \safe OS that they are willing to log in to.  Thus, for a \safe
system to release data to an untrusted device (e.g., belonging to
Alice's friend Betty), the \safe OS on the device must present a
certificate belonging to a user that is authorized to access the data
(e.g., through a \safe policy).

\subsubsection{\safe Cloud Enforcement Systems}
\safe OSes can pass sensitive user data to trusted \safe cloud
services because the OS can rely on the cloud service to respect the
user's policies.  However, we do not expect programmers to modify all
cloud services to meet the \safe guarantee, so we rely on \safe
enforcement systems to ensure that untrusted cloud services meet the
guarantee.  This paper describes two \safe cloud enforcement systems,
\emph{Magma} and \emph{Geode}.

Magma is a distributed cloud runtime for application backends;
Figure~\ref{fig:arch} shows it running the FitApp backend.  Magma
enforces the \safe requirement using fine-grained, dynamic information
flow control to track and control the flow of \safe data through
unmodified application code.  Geode, shown as the storage layer in
Figure~\ref{fig:arch}, is a storage proxy for storage systems that do
not modify application data.  It enforces the \safe requirement on
untrusted key-value stores (e.g.,
memcached~\cite{fitzpatrick04:_distr}, Redis~\cite{13:_redis}) by
interposing on storage accesses and encrypting and checksumming data.
While we believe that these systems meet the needs of many
applications, we envision the potential for other \safe enforcement
systems, including ones using existing IFC systems~\cite{giffin:hails,
  schultz13:_ifdb}, sandboxing systems~\cite{sangmin:pibox,
  lee14:_clean_approac_byoa} or computation over encrypted
data~\cite{popa14:mylar, popa11:_crypt}.

\subsubsection{\safe Verification and Attestation}
\label{sec:safe-cloud-service}
\safe enforcement systems make it easier for \safe OSes to verify that
cloud services are \safe.  Rather than simply keeping a list of \safe
cloud services, cloud services \emph{demonstrate} that they are \safe
by attesting, using trusted platform modules (TPMs), that they are
running a trusted systems stack including a \safe enforcement system.
We do not innovate here; this could be achieved using a secure
bootloader~\cite{arbaugh97:_secur_and_reliab_boots_archit,
  sailer04:_desig_implem_tcg_integ_measur_archit}, a trusted
hypervisor~\cite{garfinkel03:_terra}, or a secure enclave
mechanism~\cite{baumann14:haven}.

The trusted hardware component measures all of the software that makes
up the platform up to and including the \safe enforcement system and
produces a signed hash summarizing this software stack.  The \safe OS
validates this hash against a list of \safe software platforms.  We
imagine that these hashes could be provided by the mobile OS vendor
(much as OS and browser vendors maintain lists of trusted SSL CAs
today) or a trusted third party.  Note that it is practical for OS
vendors to distribute these hashes because we assume a limited number
of trusted system stacks and \safe enforcement systems; however, a
trusted cloud service could also do so.

As alternative (e.g., if trusted attestation hardware is not available
or a cloud vendor prefers not to reveal its deployment), the \safe
architecture can be used by having the OS vendors (or a trusted third
party) validate cloud platforms.  Then, these platforms (e.g., Amazon
Lambda~\cite{lambda} or S3~\cite{s3}) would be trusted axiomatically
to enforce the \safe property.  The OS vendor would simply issue them
a signed certificate that the cloud provider stores and presents to
mobile OS clients. Obviously, this is a less secure option than
attestation because it requires trusting the cloud providers to
correctly run a trusted \safe enforcement system and would not work
for application providers that provide their own infrastructure.
However, we offer the alternative because many applications today run
on a third-party cloud provider platform which has other strong
incentives to enforce user privacy guarantees.

\subsection{\safe Policies}
\label{sec:policies}
\safe policies are flow policies expressed as access control lists
including \emph{users}, \emph{groups}, and \emph{applications}.  A
\safe policy encodes: (1) the app that can access the data, and (2)
the list of apps, users and groups with which the app can share that
data and any data derived from it.  For example, Alice can set a
policy allowing her fitness app (FitApp) to share her GPS location
only with Betty. We use the following notation to denote this \safe
policy: $\resource{GPS} =
\tup<\app{FitApp},\set{\principal{Betty}}>$. Note that this policy
only allows FitApp to share Alice's GPS-derived data with Betty; Alice
may have different policies for other applications.

Apps create \safe groups that map to application-specific concepts and
register them with the \safe user service.  For example, FitApp could
define $\group{Alice}{Running-Group}$, which lets Alice share her runs
with her running partners. The \safe user service securely manages
these groups by querying the user through a \safe OS when an app wants
to add members to the group.  For example, if FitApp tries to add
Betty to Alice's running group, the \safe user services will request
that Alice's \safe OS accept or deny this request.

\subsection{Trust and Threat Model}
The \safe framework makes it possible for users to create a chain of
trust from their mobile devices and OSes to the cloud. Thus, we begin
with the assumption that users trust their own mobile devices and
their \safe OS.  We establish user trust by requiring users log in to
their \safe OS. This login establishes trust in several ways: (1) it
authenticates the user to the \safe OS, allowing the OS to give her
access to the data and sharing policies on the mobile device and (2)
it indicates to the \safe OS that the user trusts the device to
collect and hold her sensitive
data. 

A user login also indicates that the device and \safe OS is trusted to
hold data shared with that user.  For example, if Alice gives Betty
access to her GPS location, then she must trust any device and \safe
OS that Betty is willing to log in to.  This extension of trust makes
sense; even if Betty's phone was not running malicious software, Betty
herself could extract Alice's GPS location from the phone once it is
shared because Betty has physical ownership of the phone.

Users also trust attested \safe cloud services.  In particular, users
trust \safe services running on a trusted systems stack including a
\safe enforcement system.  Similar to other security systems, the
degree of protection offered by each \safe enforcement system depends
on its mechanism.  In general, \safe systems can suffer from timing
attacks, probabilistic channels, or physical attacks on trusted
components.  For example, Magma uses IFC to enforce \safe policies on
untrusted applications, so it suffers from many of the same
limitations as previous IFC systems~\cite{denning77:_certif,
  myers:jif, efstathopoulos:asbestos, krohn07:_infor_os,
  cheng12:aeolus}, including termination.  Despite these limitations,
the \safe framework significantly improves the security of user data
handled by distributed mobile apps.  Since, today, users must trust
their applications, \safe enforcement systems significantly raise the
bar on attacks by malicious applications on user privacy.




%% file: os.tex

Agate is a \safe enforcement operating system built atop Android, the
most popular mobile OS today~\cite{18:_global_os}; however, Agate's
design could layer atop other mobile OSes as well (e.g., iOS).  A
\safe mobile OS performs three important functions: (1) providing
users with a secure user interface for logging in and setting and
managing policies, (2) labeling data with \safe policies and (3)
enforcing \safe policies.
\subsection{Agate Architecture}
\begin{figure}[th]
  \centering
  \addfigure{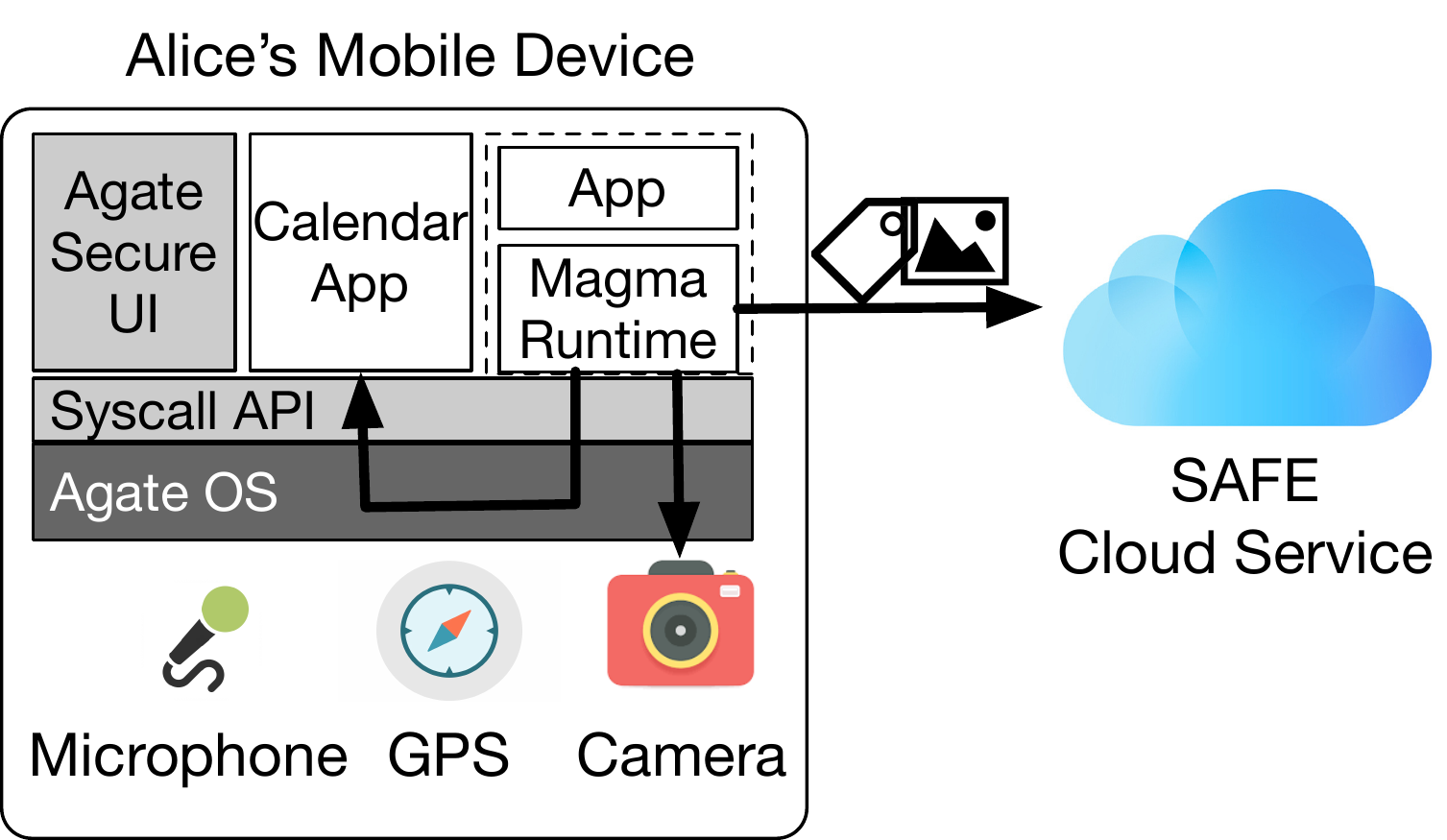}
  \caption{\textbf{Agate Mobile OS Architecture.} Agate mediates
    application access to hardware and software resources (similar to
    Android).  It provides a secure user interface for users to log in
    and set Agate policies, and a syscall interface for apps to access
    resources and propose policies.  Agate runs every app in a Magma
    runtime for \safe enforcement.  Magma ensures that apps only send
    data from OS-protected resources to other \safe systems or users
    within the policy.}
  \label{fig:agate-arch}
\end{figure}

Figure~\ref{fig:agate-arch} shows the Agate mobile OS architecture.
To minimize the impact on the user experience, we limit Agate to enforcing
\safe policies on data that mobile OSes already protect (essentially
anything in the Android Manifest).  We define these data sources as
\emph{OS-protected resources}, including hardware resources like the
camera and GPS (shown below Agate in
Figure~\ref{fig:agate-arch}), as well as software resources
provided by built-in apps, like the user's calendar (shown in the
middle of Figure~\ref{fig:agate-arch}), contacts, etc.

Agate provides two interfaces (shown in gray in
Figure~\ref{fig:agate-arch}): (1) the \emph{user interface} lets users
securely log in, set policies and manage group membership and (2) the
\emph{syscall interface} lets applications access OS-protected
resources, suggest policies and create groups.

Agate is a \safe enforcement system.  Once an app has accessed an
OS-protected resource, Agate must ensure that untrusted apps do not
send data from those resources to untrusted cloud services or mobile
OSes.  For \safe enforcement, Agate embeds the Magma runtime and runs
every app with it.  Unlike the Magma distributed cloud runtime, Agate
does not not require attestation because it is sufficient that the
logged-in user trusts Agate and its apps to be \safe.

\subsection{Agate Policies}
Agate policies match existing user privacy policies as closely as
possible.  They combine today's access control policies with \safe
flow policies and let users express: (1) which OS-protected resources
an application can access, and (2) how the application can export data
derived from that resource.  For example, Alice can set a policy
allowing her FitApp to access her GPS and share with Betty.  Agate
will give every piece of data that FitApp derives from her GPS the
\safe policy,
$\resource{GPS} = \tup<\app{FitApp},\set{\principal{Betty}}>$.


Agate lets apps create \safe groups and suggest policies through the
syscall API.  For example, once FitApp has created the group,
$\group{Alice}{Running-Group}$, then it can offer Alice the choices:
$\resource{GPS} = \tup<\app{FitApp}, \set{\principal{Alice}}>$
$\resource{GPS} = \tup<\app{FitApp},
\set{\principal{Alice},\principal{Betty}}>$ and
$\resource{GPS} = \tup<\app{FitApp}, \set{\principal{Alice},
  \group{Alice}{running-group}}>$ to share data from her GPS only with
her devices, with her and Betty's device, or with her entire running group.


\subsection{Agate Syscall Interface}
Mobile apps interact with Agate through the syscall interface.
Syscalls fall into three categories: (1) access to OS-protected
resources, (2) Agate policy proposals, and (3) \safe group management.
Most of these syscalls are directly handled by Agate or coordinated
with the \safe user service. Whenever possible, we maintain the
existing OS interface; for example, application access to OS-protected
resources is unchanged.

Existing apps access hardware resources (e.g., the camera) through
syscalls handled by the OS and software OS-protected resources through
inter-process calls to built-in OS apps (similar to a user-level file
system on a traditional OS). For example, on Android, an app can
access a user's contacts by sending an intent to the Contacts
app. When an app accesses an OS-protected resource, \agate continues
to enforce an access control policy similar to today's mobile OSes
(e.g., Alice gives \app{FitApp} access to her contacts through the
Android manifest) but labels any data from the resource (e.g., Betty's
address) with a \safe policy.  It then gives that \safe policy to
Magma to enforce on the untrusted mobile app and pass on to \safe
cloud services.

\subsection{Agate User Interface}
\marnote{Figure out how to authenticate the app}
Before the user can access \safe apps on Agate, they must log in.
Agate presents a log-in interface similar to existing mobile OSes or
apps.  It authenticates the user's identity with the \safe user
service to obtain a user certificate.  Agate can support apps from
more than one \safe ecosystem; however, users have to log in to each
ecosystem separately.  Apps provide the location of their ecosystem's user service,
so that Agate can retrieve the app id and certificate. Agate will ask
the user to log in again only if it does not already have a
certificate from that user service.

\agate requires a secure way for users to specify policies for their
OS-protected resources and manage groups.  To avoid application
interference, \agate cannot trust the application to display or draw
the policy-creation UI.  Instead, it displays the UI in a
\textit{secure user interface}, similar to the UI used today when
mobile apps request additional app
permissions~\cite{apple:_human_inter_guidel}.  Any policy-creation UI
should be secure from: (1) \textit{visual manipulation} by the
application (e.g., changing what the user sees); (2) \textit{input
  forgery} by the application (e.g., entering a policy on Alice's
behalf); and (3) \textit{clickjacking} or similar
attacks~\cite{huang12:clickjacking}.  Prior work has considered these
and other secure UI requirements in
depth~\cite{roesner13:layercake,roesner12:uist}; we refer to that work
for implementation details for these properties.

To extend \agate's support for text-based applications, we added a new
secure text box. For example, a chat application can open a secure
\agate text box, which reads text from the user and then labels it
with a policy before handing it back to the application.  We assume
the mobile OS allows users to verify that they are operating in the
context of a trusted built-in application or \agate UI (e.g., an
indicator in the system navigation
bar~\cite{bianchi15:_ui_deception}).

\subsection{Agate \safe Enforcement}
Agate performs \safe enforcement because we want to let untrusted
mobile apps manipulate user data from OS-protected resources but not
let apps leak that data to un-\safe cloud services or unattested OSes.
It leverages Magma for this enforcement by running every mobile app in
a Magma runtime environment and giving Magma \safe policies for any
data from OS-protected resources.  Magma is a modified IFC JVM, making
it suitable for mobile apps as well as cloud backends.  We leave the
discussion of how Magma implements \safe enforcement to the next
section.





%% file: platform.tex

Magma is a runtime system that provides \safe enforcement for
unmodified application processes.  Because Magma is a \safe
enforcement system, it can always pass data to another process running
the Magma runtime.  Thus, it is easy to construct a distributed
runtime platform from processes running Magma across distributed
nodes.

Magma's responsibilities as a \safe enforcement system is to: (1) take
\safe policies and turn them into IFC tags for its tracking mechanism,
(2) propagate those tags as the application manipulates user data, and
(3) check tags to ensure that \safe policies are enforced when the
application backend sends data to mobile devices.

\begin{figure}[thp]
  \centering
  \addfigure{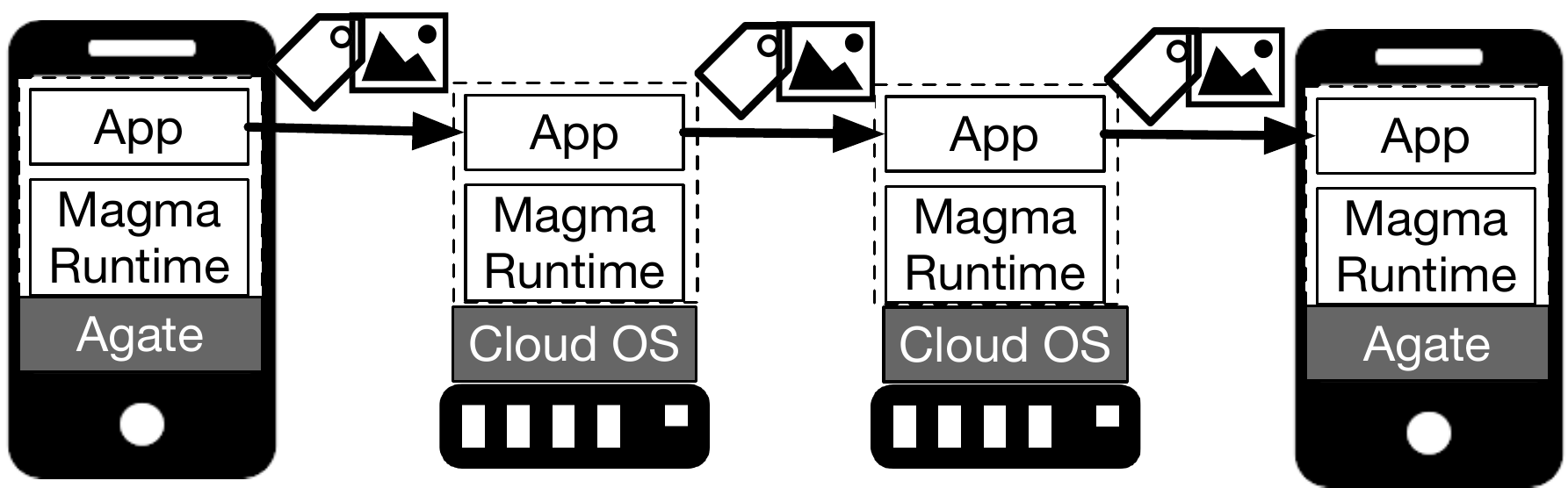}
  \caption{\textbf{Magma Architecture.} Every Magma process runs
    application code atop the Magma runtime and can always pass data
    to another (verified) Magma process.  Magma processes can run
    embedded in Agate OSes on mobile devices or as stand-alone
    components on cloud servers.}
  \label{fig:magma-arch}
\end{figure}

\subsection{Magma Architecture}

Figure~\ref{fig:magma-arch} shows the architecture of a distributed
Magma runtime system.  Every Magma process (shown as dotted
rectangles) runs the Magma application runtime under the application.
Magma processes can run on mobile devices as part of the Agate OS or
as part of a cloud service.

We prototyped the Magma runtime by extending the Dalvik JVM to support
Android apps on Agate.  However, other application runtimes (e.g.,
Python, Scala, Go) could be used as well.  Using a language runtime
lets Magma support fine-grained IFC with low overhead.  In contrast,
supporting low-level languages would cause Magma either to overtaint
or impose too much performance overhead. We observe that most sharing
applications today run in a managed runtime, so this trade-off is a
reasonable way to achieve our goal of supporting unmodified
applications.

Magma uses fine-grained information tracking, rather than tainting
processes, because backend sharing app code may handle the data of
many users over time.  Eventually, processes running the cloud service
would become so tainted that they would not be able to release data to
any users.  Another option would be to start a new process for each
request (e.g., using something like Amazon Lambda~\cite{lambda});
however, that could add significant latency, so we leave that option
to future work.

\subsection{Magma IFC Model}
\label{sec:magma-ifc-model}
Magma's IFC model is similar to other IFC systems with one key
difference: \emph{\safe policies directly map to IFC labels.}  As a
result, Magma is able to work with unmodified applications without the
help of programmer or users.  Magma automatically tags data from an
Agate OS or another \safe cloud service with the \safe policy before
handing that data to the application.  Magma propagates and enforces
those policies represented as tags across the entire backend cloud
application.

There are three types of IFC principals in Magma -- users, groups and
applications -- which map to the principals in \safe flow policies.
We directly use \safe ids for Magma labels.  There are two types of
labels in Magma: the mutable \emph{data labels}, which carry the
\safe policy, and the immutable \emph{process labels}, which encode
authorization and help enforce the \safe policies.

\paragraph{Data Labels.}
Magma data labels are tags of the form $l
=\set{\principal{o$_1$, o$_2$} \to \principal{a$_1$, u$_1$,
    u$_2$, g$_1$}}$, where \principal{o$_1$, o$_2$} are the user
principal ids of the owners of the labeled data, $\principal{a$_1$}$
is the principal id of an application allowed to access the labeled
data and $\principal{u$_1$}$, $\principal{u$_2$}$, $\principal{g$_1$}$
are users and groups allowed to view the labeled data. We will refer
to the set $\{$\principal{a$_1$, u$_1$, u$_2$}, $\forall$ user
principal id $u \in$ \principal{g$_1$}$\}$ as $readers(l)$. 

\paragraph{Process Labels.}
Each Magma process is labeled with an application principal (e.g.,
$\{\principal{a}_1\}$) and, if the process runs on Agate, a user
principal (e.g., $\{\principal{a}_1, \principal{u}_1\}$).  Magma
requires these two to enforce the \safe flow policies, which dictate
which application can move a user's data, as well as, whom the app can
give that data to.

\paragraph{Information Flow Rules.}

To enforce \safe policies, Magma must guarantee the following security
property:
\begin{itemize}
\item[] \emph{Data from an OS-protected resource labeled with a
    non-empty initial data label $l_1$ may reach a process labeled
    with label $l_2$ only if $l_2 \subseteq readers(l_1)$}.
\end{itemize}
It does so by applying the following two data and policy propagation
rules:
\squishlist
\item[] \textbf{Intra-process propagation.} Inside a process, data is
  allowed to flow freely (i.e., no flow control checks are performed)
  but data labels may change. Any data \emph{derived}
  from one or more labeled data sources is labeled with a label which
  reflects all the \safe policies involved. For example, if the
  application combines two pieces of labeled data, their labels are
  \emph{merged} into the resulting label $l$, where the owners are the
  union of the two sets of owners and where $readers(l)$ is the
  intersection of the two sets of readers. That is, given two pieces
  of data with labels $l_1 = \set{\principal{o$_1$} \to readers(l_1)}$
  and $l_2 = \set{\principal{o$_2$} \to readers(l_2)}$, the resulting
  label of any derived data is $\set{\principal{o$_1$,
      o$_2$} \to readers(l_1) \cap readers(l_2)}$.
\item[] \textbf{Inter-process propagation.} Data labeled with the
  current data label $l_1$ is allowed to flow to a process labeled
  with label $l_2$ only if $l_2 \subseteq readers(l_1)$. If the data
  is allowed to flow to the new process, it maintains its label,
  $l_1$, until an intra-process propagation rule is applied.
\squishend

These rules ensure that data protected by \safe policies, and data
derived from that data, flows only to other processes running the same
app (or another allowed app), and, if the process is running on an
untrusted Agate OS, a process with a logged in user that is in the
\safe policy.

\subsection{Magma Flow Tracking}
Magma implements both explicit and implicit flow tracking.  Explicit
flows are caused by direct assignment (e.g., \code{x = gps-loc}),
whereas implicit flows are caused by control flow (e.g., \code{if
  (gps-loc == home) \{x = true\}}.  Magma's explicit flow tracking
mechanism is relatively straightforward; as apps propagate data, Magma
propagates the corresponding tags, joining the tags by following the
IFC rules.  Handling implicit flow through unmodified applications is
more complicated, so much so that many systems
\cite{enck10:taintdroid, rosen13:_apppr, giffin:hails} ignore the
problem altogether.  However, they are an important way that user
privacy can be violated; thus, Magma must consider them.

Using the previous example, \code{if (gps-loc == home) \{x = true\}},
the value assigned to \code{x} is a literal value containing no
sensitive labels.  However, the execution of the assignment operation
reveals information regarding Alice's location.  It is also worth
noting that information is leaked even if the conditional branch is
\emph{not} executed since the absence of an update to \code{x} also
reveals information regarding Alice's location (i.e., she is not
home).

For implicit flow tracking, Magma uses a combination of static
analysis and runtime taint propagation.  For every conditional block,
Magma's static analyzer identifies both the set of variables that are
updated in either branch (e.g., \code{x}) and the control flow
variables that determine the conditional execution of the updates
(e.g., \code{gps-loc}). The static analysis pass then inserts code
that causes the runtime taint propagation system to update the labels
of the modified variables to include the labels associated with the
control flow variables, regardless of whether the conditional is
executed.

Our Magma prototype runs this static analysis on Java bytecode as it
loads apps. It uses a control flow graph representation of the program
and resembles the techniques outlined in
\cite{clause07:_dytan,Kang11_DTA,cox14:_spand}. Instrumented code for
runtime taint tracking needs to be added to every control flow block
that might contain sensitive data, as well as any function invoked
from either branch of these control flow blocks.

Instrumenting control flow has the potential to increase code size and
execution time. To mitigate this problem, Magma's static analyzer
makes a conservative pass to determine which control flow blocks will
\emph{never} result in implicit flows, because they never access
tainted data. Similarly, it identifies functions that are never called
from tainted contexts.  We found this pruning to be useful in practice
because many tagged data objects are unlikely to be used to make
control flow decisions: for example, most apps do not branch on the
bytes of a JPEG image.

\subsection{Magma \safe Enforcement}
Magma uses flow control to enforce \safe policies on untrusted
applications.  Because Magma directly expresses \safe policies as IFC
labels, it can use labels tagged on the data to check policies.  Thus,
simply by preventing flows that violate Magma's IFC rules, Magma can
ensure that an application meets the \safe requirement.

First, Magma always permits applications to send data to another
trusted \safe cloud service.  Magma verifies that cloud services are
safe using one of the methods detailed in
Section~\ref{sec:safe-cloud-service} (e.g., comparing a TPM hash).  It
then translates the IFC label into a \safe policy and securely send it
along with the data.

When a Magma application sends labeled data to an unattested mobile
device (i.e., belonging to another user), Magma performs a policy
enforcement check.  It first checks that the mobile device is running
an authorized \safe OS and retrieves the application id and logged-in
user id of the process that will receive the data.  The user id and
app id are both provided as signed certificates from the \safe user
service when the user logs in and starts the app.

If the app id matches, then Magma intersects $readers(l)$, where $l$
is the label on the data being sent, with the logged-in user principal
of the destination process.  If the intersection is not empty, then
the Magma tags are translated into a \safe policy and sent together to
the destination.  If the intersection is empty, then the destination
is not permitted to receive the data and Magma returns an error to the
application.  This check is sufficient to ensure \safe policies are
respected.

%% file: storage.tex

Application cloud back-ends often need to persistently store data for
fault-tolerance or archival storage. To support this requirement, we
provide a cryptographic proxy, Geode, to make existing, untrusted
storage systems \safe.  Geode takes its approach to building secure
storage from untrusted infrastructure similar to prior work on
TPM-based filesystems and
databases~\cite{maheshwari00:_how_to_build_trust_datab,
  weinhold08:_vpfs, chen08:_overs}.

\subsection{Geode Interface and Guarantees}
Geode provides a key-value object storage interface, like Amazon
S3~\cite{s3}. It provides three guarantees: (1) \emph{confidentiality}
- the storage service cannot read user data, and it will be released
only according to the \safe policy on the data; (2) \emph{integrity} -
each object and its policy can only be modified by the application
that created it and cannot be tampered with by the storage service,
and (3) \emph{single-object linearizability} of updates.

Geode provides linearizability per object both because it is a
desirable property for reasoning about concurrency, and to prevent a
malicious storage service from returning incorrect values under the
guise of weak consistency.  Geode can be used with any storage service
that provides the same interface, guarantees linearizability, and does
not manipulate user data itself.  Many storage systems meet these
requirements, including most weak consistency distributed storage
systems (e.g., S3, Redis~\cite{13:_redis}), which typically provide
per-key linearizability.

\subsection{Geode Architecture}
\begin{figure}[th]
  \centering
  \addfigure{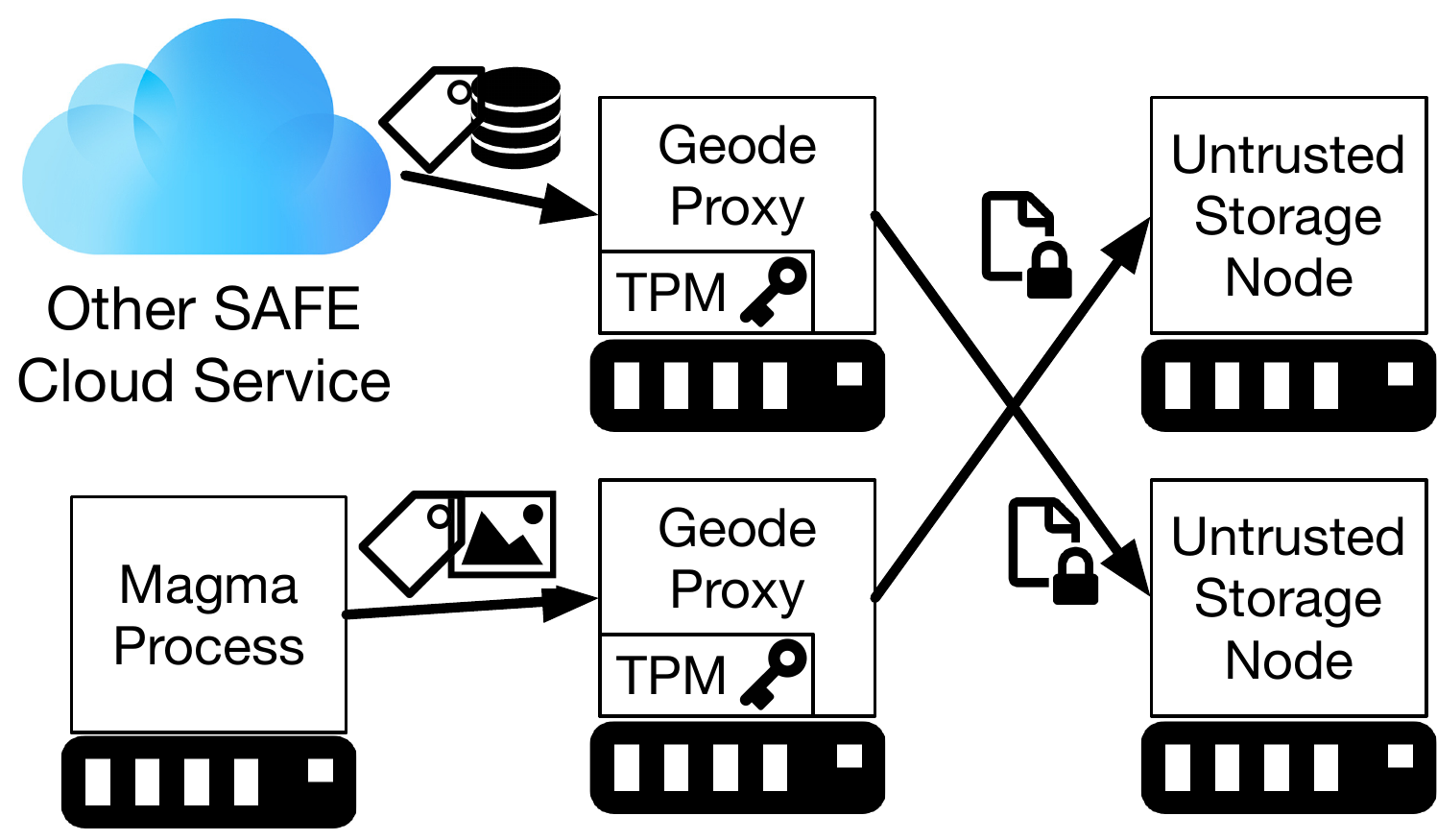}
  \caption{\textbf{Geode Architecture.} Geode interposes on access to
    an untrusted storage system from \safe systems.  It securely
    checksums and encrypts the data and \safe policies before handing
    them to the storage system.  Geode ensures that only \safe systems
    and users within the \safe policy can retrieve the data.}
  \label{fig:geode-arch}
\end{figure}

Figure~\ref{fig:geode-arch} shows Geode's architecture.  Geode
operates multiple proxy nodes, each responsible for a different
portion of the keyspace. Due to space constraints, we do not discuss
fault-tolerance of proxy nodes, other than to note that it can be
handled by replication and logging techniques as in previous
systems~\cite{maheshwari00:_how_to_build_trust_datab,
  weinhold11:_jvpfs}. Each Geode node is equipped with a TPM or other
trusted hardware component; in addition to attesting to \safe clients
that the server is running the Geode proxy, it also provides the Geode
proxy with access to a sealed encryption key and a tamper-proof
monotonic counter.

Note that Geode could be the only cloud service that an app needs.
Many applications today (e.g., to-do lists or recipe apps) use only a cloud
storage system (e.g., Dropbox~\cite{15:_dropb}) and do not require a
separate cloud backend app.

Geode could be deployed in a number of ways.  It could be run by the
application provider, the storage provider or a third party entity.
For the best performance, Geode should be co-located with the storage
system. 

\subsection{Geode \safe Enforcement}
Geode interposes on every access to the storage system. For each write
operation, Geode prepends a header to every stored object with its
\safe policy. It then generates a random initialization vector and
uses it to encrypt the block (using AES-128 in CBC mode with the
secure key). Encryption ensures that the storage system cannot read
the data, and the initialization vector prevents known-plaintext
attacks. Subsequently, it records the initialization vector and a
SHA-256 HMAC of the block contents. These are added to a per-object
integrity table, itself encrypted with the same key, which also
contains the latest monotonic counter value; this table is stored in a
special object in the underlying storage system. (A more efficient
implementation might use a Merkle
tree~\cite{merkle79:_secrec_authen_and_public_key_system} to prevent
having to rewrite the table on each update.)

On each read operation, the Geode proxy fetches and decrypts the
object. It verifies that the hash of the object matches the one stored
in the integrity table, and the version number of the table is up to
date. The hash ensures that the storage service did not tamper with
the object or its policy, and the version number prevents it from
rolling back to an earlier state. If the data object has a \safe
policy, Geode first checks if the requesting application is either a
\safe service or a \safe OS. If it is a \safe service, Geode securely
sends the decrypted data and \safe policies to the service. If it is a
\safe OS, Geode checks the \safe policy for the user certificate
presented by the \safe OS (similar to Agate and Magma).

%% file: eval.tex

In addition to ensuring that user privacy policies are respected, we
stated three goals in the \safe framework design: (1) minimize changes
to the user experience, (2) minimize changes to application code and
(3) minimize performance cost.  In this section, we evaluate the
effectiveness of the \safe design in meeting these goals.
 
\subsection{Implementation}
To support Android sharing apps, we prototyped Agate and Magma using
the Android OS and Dalvik JVM, respectively.  Agate runs on ARM-based
mobile devices while Magma runs on both ARM and x86 architectures.
This section describes the implementation of these prototypes.

\subsubsection{Agate Mobile OS Prototype}
Agate extends Android into a \safe OS by adding support for secure
user log-in, policy collection and \safe enforcement.  Our prototype
UI differs slightly from the one imagined for Agate; because Android
already provides users with a secure way to set access policies for
their OS-protected resources, the Agate UI only provides settings for
\safe policies. Our policy management interface consists of dialog
boxes drawn in the context of the current application rather than in a
separate, trusted application.  A more secure prototype would show the
policy interface in a separate application, as demonstrated in prior
work~\cite{roesner13:layercake}.

Our current prototype leaves access control to existing Android
mechanisms but interposes on accesses to collect policies and attach
labels for Magma to use.  After attaching labels, Agate uses its
embedded Magma runtime to enforce the \safe guarantee. Our prototype
interposes only on calls to the built-in camera and GPS resources
(i.e., the \texttt{takePhoto()} and \texttt{getLastKnownLocation()}
system calls); in a full implementation, similar modifications would
be required for other OS-protected resources.

\subsubsection{Magma Prototype}
Magma is a \safe enforcement runtime that: (1) translates \safe
policies into IFC tags, (2) tracks both explicit and implicit flows and
(3) enforces \safe policies with IFC checks.  Magma's explicit flow
tracking mechanism is based on TaintDroid~\cite{enck10:taintdroid} for
Android 4.3\_r1.  However, TaintDroid is a limited taint-tracking --
not flow-enforcement -- system, so Magma requires extensive
modifications to TaintDroid's mechanisms.  For example, while
TaintDroid tracks binary taints for only 32 sources, Magma must use
more complex IFC labels and rules to represent \safe policies.

TaintDroid has no implicit flow tracking, so Magma implements its own
mechanism.  Magma's hybrid mechanism uses a custom analysis tool to
insert annotations into Android dex files, and then dynamically
propagates labels at runtime via those annotations. Magma's static
analysis tool uses the Soot framework for Java/Android
apps~\cite{soot}, and consists of 5,400 lines of Java code to perform
class hierarchy analysis, global method call flow analysis, control
flow analysis within methods, side effect analysis inside
conditionals, and insertion of taint tracking code for implicit flows.

Magma inherits some performance optimizations from TaintDroid that
could lead to overtainting. Neither TaintDroid nor Magma performs
fine-grained flow tracking through native code due to the performance
overhead. Instead, Magma uses a conservative heuristic that assigns
the result of a native code function to a combination of the taints of
the input arguments.  Similarly, TaintDroid keeps a single taint label
for an entire array, which could cause overtainting due to false
sharing.  Phosphor~\cite{bell14:phosphor}, a newer JVM-based taint
tracking mechanism, eliminates the potential for overtainting at a
reasonable performance cost.

\begin{table}[t]
  \centering
  
  \caption{\textbf{Protection offered by Android and the \safe
      framework against top web and mobile
      vulnerabilities~\cite{OWASP-vulnerabilities,owasp-mobile}.}
    Related items from the lists are merged. Android handles only a
    small subset of these issues, while \safe covers nearly all.}
  
  {\small
    \def\arraystretch{1.0}
    \def\cellset{\def\arraystretch{1}}
    \def\theadfont{\bfseries\footnotesize}
    \def\theadalign{ll}
    \def\theadset{\def\arraystretch{1.3}}%
    \def\cellalign{l}
\scalebox{0.94}{
    \begin{tabular}{@{}lcc@{}}
    \toprule
    \thead{Vulnerability}&\thead{Android}&\thead{\safe}\\
    \midrule
    Broken access control & -- & \checkmark \\
    Broken authentication & -- & \checkmark \\
    Broken cryptography & -- &\checkmark \\
    Buffer overflow & \checkmark & \checkmark\\
    Client or server side code injection & -- & \checkmark \\
    Cross site scripting & \checkmark & \checkmark \\
    Insecure data storage & -- & -- \\
    Insecure direct object references &\checkmark &\checkmark \\
    Insufficient transport layer protection & -- &\checkmark \\
    Improper error handling & -- & \checkmark \\
    Improper session handling & -- & \checkmark \\
    Lack of binary protections & -- &\checkmark \\
    Missing function-level access control & -- &\checkmark \\
    Path traversal \& command injection (server) & -- &\checkmark \\
    Security decisions via untrusted inputs & -- &\checkmark \\
    Sensitive data exposure & -- &\checkmark \\
    Unintended data leakage & -- &\checkmark \\
    \bottomrule
  \end{tabular}}
  }
  \label{tab:vulnerable}

\end{table}

\subsection{Security Analysis}
We first evaluate the effectiveness of \safe{}'s security guarantees.
We examine the top 10 web security risks and the top 10 mobile
security risks identified by the Open Web Application Security Project
(OWASP)~\cite{OWASP-vulnerabilities,owasp-mobile}, summarized in
Table~\ref{tab:vulnerable}. As the checkmarks in
Table~\ref{tab:vulnerable} indicate, the \safe framework can
successfully prevent applications from leaking user data for nearly
all of the most common web and mobile vulnerabilities compared to
Android, which handles only three.

In many cases, simply using a \safe enforcement system suffices to
avoid the vulnerability. For example, applications frequently
inadvertently leak user data and violate user policies through {\em
  improper error handling}.  However, an IFC-based enforcement system,
like Magma, would ensure that applications cannot release user data
except to another \safe system or a user within the \safe policy
running a \safe OS.  In fact, while porting applications to Agate, we
often found Magma barring the release of error message for debugging
to us.

Finally, the entire \safe framework is designed to prevent {\em broken
  access control}.  Rather than trusting applications to correctly
implement access control checks, a \safe OS requires that a cloud
service run a \safe enforcement system, which is trusted to perform
these checks on behalf of the application.  For example, Magma would
not allow Facebook to release Mark Zuckerberg's photos to users that
are not in his friends
list~\cite{soze11:_mark_zucker_privat_photos_leaked}.

While the \safe framework protects against \emph{insecure data
  storage} in the cloud (i.e., by using the Geode proxy), we
explicitly do not handle risks that can be exploited by improper
storage on untrusted user devices.  For example, once Alice releases
her photo to Betty, she must trust that Betty does not copy the photo
to untrusted cloud storage or lose her phone.  Overall, the analysis
shown in Table~\ref{tab:vulnerable} demonstrates that, through the use
of the \safe framework, taking trust away from applications and
putting it into the hands of attested \safe enforcement systems can
help users avoid many of the vulnerabilities that plague mobile
sharing applications today.

\begin{figure}[tb]
\begin{center}
\begin{minipage}{0.22\textwidth}
\includegraphics[width=1.55in,height=2.3in]{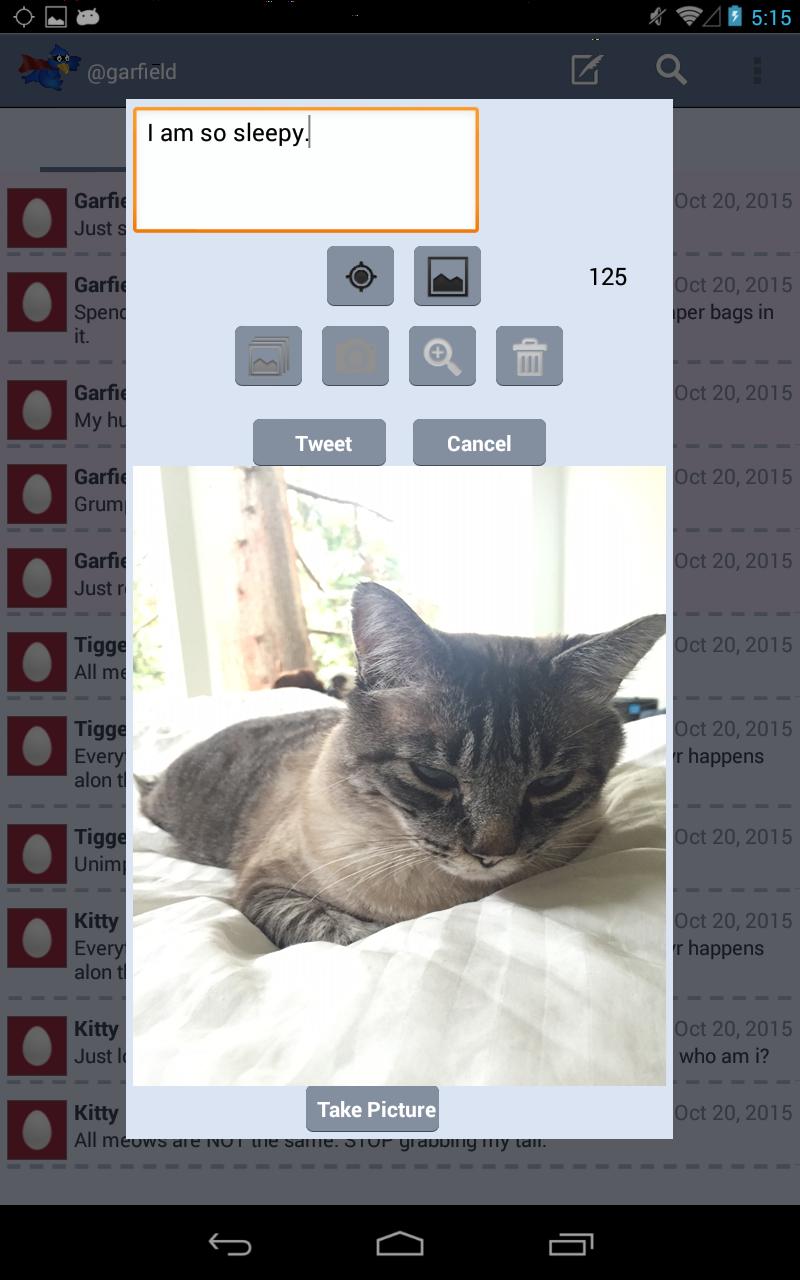}
\subcaption{}
\end{minipage}\hspace{0.1in}
\begin{minipage}{0.22\textwidth}
\includegraphics[width=1.55in,height=2.3in]{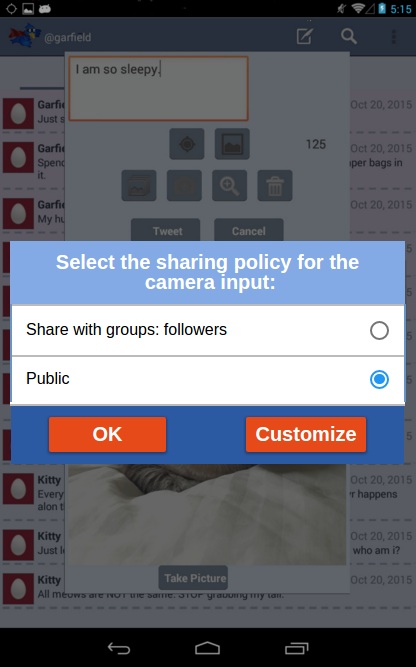}
\subcaption{}
\end{minipage}
\vspace{-.2in}
\end{center}
\caption{\textbf{Agate UI for a Twitter-like application.} When the
  user takes a photo in \safe{}Tweet, shown in 5(a), Agate interposes
  on the system call and lets the user set a policy via a secure,
  system-controlled user interface, shown in 5(b).  Agate labels the
  photo with the specified \safe policy, which Magma enforces, once
  the photo is given to \safe{}Tweet.}
\label{fig:ui}
\end{figure}

\subsection{User Experience}
A key goal of the \safe framework is to minimize changes to the user
experience and use existing user policies.  We evaluate whether we
achieved this goal by exploring the Agate user interface.

To demonstrate Agate's UI, we use an open-source Twitter
clone~\cite{15:_twimight} that we ported to our \safe framework, which
we call \safe{}Tweet.  While interacting with \safe{}Tweet, Alice
takes a photo of her cat, Max, to post to her feed, which requires
\safe{}Tweet to access her camera.  At this point, Agate interposes on
the system call to: (1) ask permission for \safe{}Tweet to access the
camera and (2) let Alice to set a \safe policy for all data that
\safe{}Tweet receives from the camera (i.e., photos).  Once Alice has
given permission and set a policy, Agate will give the photo and \safe
policy to the embedded Magma runtime to return to the app.

We note two key aspects of Agate's UI that are enabled by the \safe
framework.  First, it looks and feels much like a user's experience
with Twitter because \safe{}Tweet can propose \safe policies that match
those that it would offer today.  Second, although \agate creates a
familiar user experience, it can enforce \safe policies on untrusted
apps using its Magma runtime \emph{and} verify that cloud services
will also enforce those policies.  So while the user experience
remains unchanged, the security properties are completely different
with the \safe framework.

\subsection{Programmability and Porting Experience}
Another goal of the \safe framework is to minimize application code
changes.  To gain experience with \safe applications, we created three
\safe sharing applications.  We ported three unmodified Java server
applications as cloud app backends and three Android apps as mobile
app clients, as listed in Table~\ref{tab:apps}.  Using the Agate UI,
we placed \safe policies on different sources of data for each
application; for example, in \safe{}Chat (70K LoC), we used \agate's
secure text input facility to create a private chat between Alice and
Betty.

To port mobile app clients to Agate and Magma, the only changes made
to the application code were those needed to incorporate \agate APIs
(e.g., to use the system call to propose policies and treat invalid
flow exceptions).  In \safe{}Cal (250K LoC), we found both explicit and
implicit flows that might violate Alice's policy; e.g., Bob, who is
not Alice's co-worker, cannot view Alice's meetings this week (an
explicit flow), or check whether Alice is free at 3 on Tuesday (an
implicit flow).

With Magma, we were particularly interested in issues with
overtainting, policy accumulation on data, and unexpected flow
restrictions.  We experienced no problems with the application code
itself: indeed, the cloud app backends (Openfire, MinnieTwitter,
Calendar) required \emph{no modification} to run on Magma.  However,
we did encounter overtainting in libraries used by cloud backend apps,
particularly for communications and parsing, such as the core Java
libraries (BufferedReader, OutputStreamReader), Java RMI, and dom4j.
For example, the message serialization libraries reuse memory buffers,
leading to unnecessary data overtainting and policy accumulation and
blocking valid flows.  After manually fixing the problematic
libraries, the applications worked as expected.  Note that these fixes
could be reused for other applications by releasing Magma-compatible
Java libraries.

While we believe Agate and Magma are two representative \safe systems,
the programming experience will vary between different \safe OSes and
\safe enforcement systems.  However, with a variety of options,
programmers can choose the best one for their cloud backend or mobile app.

\begin{table}[t]
  \centering
    \caption{For each distributed application, we list the
    unmodified Android app ported to \agate for the client side and the
    unmodified Java app ported to Magma for the server side, along with their
    size in lines of code.
  }
  {\small
    \scalebox{0.9}{
    \begin{tabular}{@{}lllll@{}}
    \toprule
    \thead{App}
    &\thead{Ported Client}
    &\thead{LoC}
    &\thead{Ported Server}
    &\thead{LoC}\\
    \midrule
    \safe{}Chat
    &Xabber~\cite{xabber:_xabber}
    &78K
    &Openfire~\cite{openfire:_openfire}
    &190K \\
    \safe{}Tweet
    &Twimight~\cite{15:_twimight}
    &13K
    &MinnieTwitter~\cite{15:_minnie_twitter}
    &1.2K \\
    \safe{}Cal
    &aCal~\cite{android_caldav_clien}
    &30K
    &Cosmo~\cite{1and1:_cosmo_calen_server}
    &40K \\
    \bottomrule
  \end{tabular}
    }
  \label{tab:apps}
  }
\end{table}

\subsection{Performance}
Finally, we measure the performance cost of meeting the \safe
requirement.  In our experiments, mobile devices run Agate with its
embedded Magma runtime and cloud servers run the distributed Magma
runtime.  Each server contained 2 quad-core Intel Xeon E5335 CPUs
with 8GB of DRAM running Ubuntu 12.04.
The mobile devices were first-gen Nexus 7 tablets (1.3 GHz quad-core
Cortex A9, 1~GB DRAM).  Our servers shared one top-of-rack switch and
connected to tablets via a local-area wireless
network.

\subsubsection{Microbenchmarks}
As a baseline, we compare the performance of Magma running on Agate to
unmodified Dalvik and TaintDroid running on Android.  We execute
mobile apps without tainted data. We use CaffeineMark 3.0, a
computationally intensive program commonly used as a microbenchmark
for Java.  CaffeineMark does not access data with \safe policies, so
there are no tags to be tracked; however, TaintDroid and Magma must
still propagate and merge empty labels, so this measurement gives us a
lower bound on the performance cost for each system.

\begin{figure}[t]
  \centering
  \addfigure{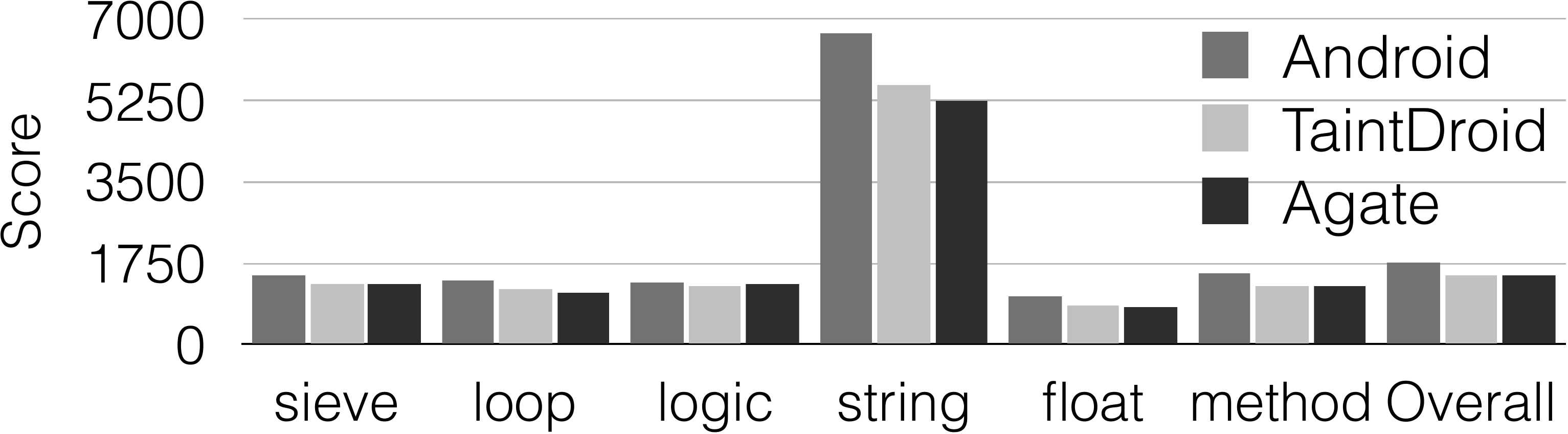}
  \caption{\textbf{CaffeineMark microbenchmark for Android Dalvik,
      TaintDroid and \agate (higher is better).} Both \agate and
    TaintDroid impose an overhead, but \agate's overhead is similar to
    TaintDroid's.}
  \label{fig:caffeine}
\end{figure}

CaffeineMark scores roughly correlated with the number of Java
instructions that the JVM interpreter executed per second. The overall
CaffeineMark score is the geometric mean of the individual scores.
Figure~\ref{fig:caffeine} shows the results for an Android tablet.  On
this baseline (no \safe policies and no tags), Agate and Magma impose
little overhead relative to TaintDroid (between .3\% and 5.8\%, with
an average of 1.8\%).  The CaffeineMark overall score for Agate is
within 16\% of baseline Android, while TaintDroid's overall score is
within 14.5\% of Android.

TaintDroid's overhead stays constant as data accumulates more taint
because it uses a single-bit representation and tracks only 15
OS-protected resources.  Magma also propagates a single 32-bit tag per
primitive or primitive array, but these tags are references to a list
of all policies with which the data is tainted.  Thus, while Magma's
cost for propagating tags remains constant as taint accumulates, the
cost of merging when combining two tainted pieces of data increases
with the number of policies tainting the data.

To evaluate the impact of accumulating taint in Magma, we measure the
cost of merging up to 20 \safe labels, each with 20 principals/tags
(10 users and 10 groups).  We consider this measurement an upper bound
because it would require the application to have at least 20 policy
options, \emph{each} with 20 principals. In practice, policies would
become unwieldy for both applications and users at a much smaller
number (i.e., 5-10 policies each with a small number of principals).
We found that merging two labels with 20 principals each took 6
\textmu s. Overhead increases with the number of labels on the data up
to 20 \textmu s to merge 20 labels with 20 tags
each. 

\subsection{Application Performance}
To validate our expectations about how apps and users use \safe
policies, and to measure \agate's performance overheads for a full
application, we use two distributed applications: (1) the \safe{}Tweet
app mentioned above (MinnieTwitter + Twimight), and (2) a multi-player
game (WordsWithFriends) that we implemented from scratch.  We run
mobile client apps with Magma on Agate and the app backends on the Magma
distributed cloud runtime.

\begin{figure}[t]
  \centering
  \addfigure{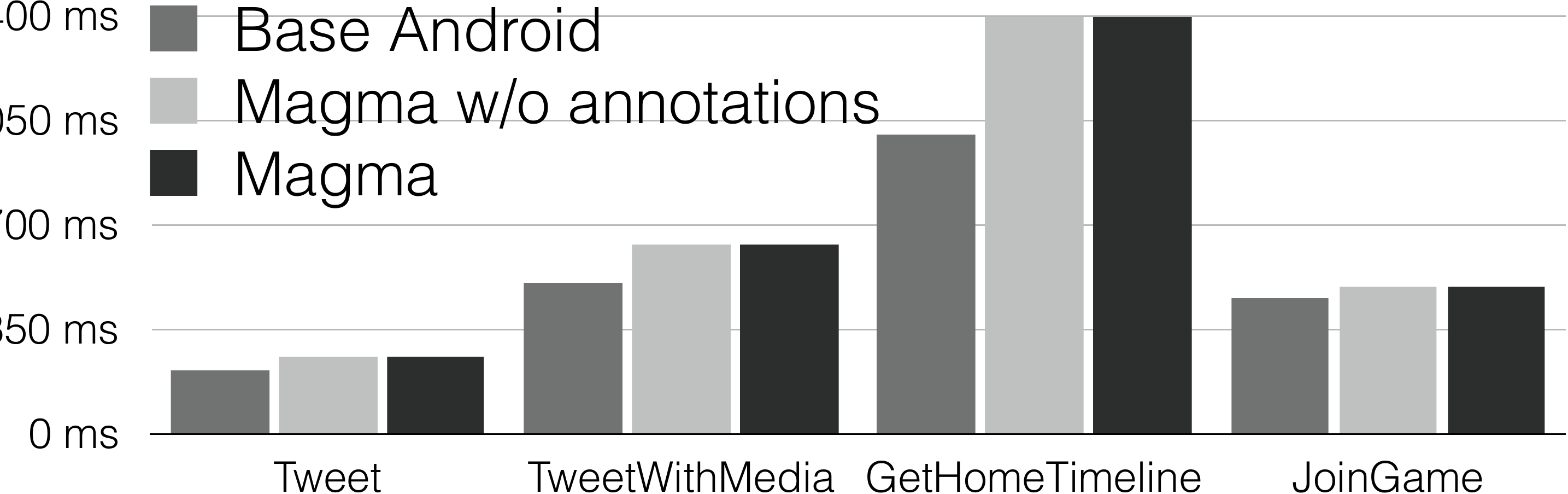}
  \caption{\textbf{Latency of \safe{}Tweet and WordsWithFriends
      functions.} The figure shows that Magma imposes roughly 20\%
    overhead compared to unsecured Android.  \textsf{joinGame} is the
    only function that requires static annotation, but those costs are
    extremely low.}
  \label{fig:twittermark}
\end{figure}

Figure~\ref{fig:twittermark} shows latency for \textsf{tweet},
\textsf{tweetWithMedia} and \textsf{getHomeTimeLine} from \safe{}Tweet,
and \textsf{joinGame} from WordsWithFriends.  For each function, we
show latency for the unsecured app on Android, Agate+Magma without
static analysis annotations, and Agate+Magma with annotations.

\textsf{Tweet} does not access sensitive data with \safe policies and
thus shows the basic cost of making Android \safe, which is 17\%.
\textsf{TweetWithMedia} includes a labeled photo; we used the default
policy suggested by \safe{}Tweet, which is
$\resource{Camera} =
\tup<\app{\safe{}Tweet},\set{\group{User}{Followers}}>$.  Overhead
increases to 20\% because Magma must call the \safe user service to
translate group/user names into principal ids.  Some optimizations
could reduce this cost, including caching id mappings and optimizing
TaintDroid's disk writes.  The overhead of \textsf{GetHomeTimeline} is
22\% because it accesses more tainted data (2 photos), requiring
\agate to check the policy on each photo. Most of the extra time is
due to remote calls to resolve group membership, which again could be
reduced with caching~\cite{cheng12:aeolus}.  Currently, each operation
required three RPCs to perform the checks before sending the photo.

For \safe{}Tweet, merging and static analysis did not add to the
runtime overhead. Magma propagates but never merges labels because
\safe{}Tweet never derives new data from photos.  Static analysis
found no implicit flows as \safe{}Tweet never uses photos as branch
conditions.  We expect this behavior to be typical for most
applications that access photos.

WordsWithFriends' \textsf{JoinGame} operation accesses the GPS to find
nearby friends for game-play, so its default policy for the GPS is
$\resource{GPS} =
\tup<\app{WordsWithFriends},\set{\group{User}{Friends}}>$.
Static analysis added only five annotations for WordsWithFriends
because the control flow based on the GPS is limited to checking for
nearby friends.  

\textsf{JoinGame} branches on tagged GPS locations every time it
compares two locations, so it must track more labeled data than
\safe{}Tweet operations and accumulates taint as it runs due to the
comparisons. However, \textsf{JoinGame} does not release the user's
GPS location (because it performs the comparisons on the user's
device), so it does not have to resolve user names or group membership
for enforcement checks, incurring a lower overhead than \safe{}Tweet
operations.

While these applications are prototypes, we believe they use the \safe
framework in a representative way. Data derived from OS-protected
resources with \safe policies were typically copied or transmitted but
rarely merged with other sensitive data.  Furthermore, static analysis
is effective at determining which conditionals operated on sensitive
data.  As a result, Magma is able to reduce the number of taint merge
operations and implicit flow annotations, and thus keep the overheads
low.

Overall, our results show that Agate and Magma's performance is very
close to TaintDroid's and within approximately 20\% of the performance
of the base Android system, both of which have no policy enforcement.
From a user's qualitative point of view, the difference is not
detectable in using either prototype application.  Although we have
not yet tried to optimize our prototype, we feel that this difference
is well worth the additional privacy guarantees that the \safe
framework provides.

%% file: relwork.tex

In designing the \safe framework we drew inspiration from many
existing privacy-preserving and trust management systems. Although
this paper presents the design of only three \safe enforcement
systems, we envision many others being built atop existing systems.

Modern OSes for smartphones incorporate access control mechanisms that
let users control which resources applications can access, e.g.,
through an Android manifest file.  Significant
research~\cite{acar16:_sok} builds on this idea to give users better
security and more effective access control.  For example, Access
Control Gadgets~\cite{franzi12:acg} provide a more intuitive user
interface for permission granting, and
Preservers~\cite{kannan11:preservers} lets users choose the
code/policies that mediate data access.  Our \safe framework and its
\safe systems provide stronger guarantees by enforcing user policies
beyond the phone.

Distributed platforms like $\pi$Box~\cite{sangmin:pibox},
Cleanroom~\cite{lee14:_clean_approac_byoa} and
Radiatus~\cite{cheng16:_radiat} protect user privacy by isolating each
user in a sandbox environment.  While isolating users makes it easy to
enforce privacy for applications where users do not interact, social
applications where users \emph{want} to selectively share their data
do not work well. All communication between users must go through a
restrictive interface and be vetted by the system. However, sandboxing
has the advantage that all data (not just OS-protected resources) is
protected and cannot be exposed even to the application
developer. $\pi$Box uses differential
privacy~\cite{dwork06:_differ_privac} to control \emph{how much} data
can be released to the developer. A sandboxing-based \safe enforcement
system would be useful for apps in which part of or all of a user's
data does not need to be shared with other users; for example, a cloud
service that categorizes a users photos using machine learning.

Unlike other IFC runtimes~\cite{giffin:hails, cheng12:aeolus,
vachharajani04:rifle, wang19:_riverbed}, Magma is explicitly designed to
support \safe flow policies.  As a result, Magma can directly translate \safe
flow policies into IFC labels, rather than requiring users or programmers to
set IFC policies.  This design minimizes changes to both the user interface and
application code.

However, the \safe framework makes it possible to incorporate other
IFC-based systems, provided that there is a way automatically
translate the \safe policies into their IFC policy model.  This
requirement may limit the untrusted applications they can support.
For example, IFC-based systems that use coarse-grained
tracking~\cite{efstathopoulos:asbestos, zeldovich:histar,
  krohn07:_infor_os} could offer better performance than Magma but
require more information about the application's architecture to deal
with overtainting.  Language-based IFC systems~\cite{myers:jif,
  liu:fabric, roy09:laminar, yip:data_flow} seem even less suitable
because they require information about application variables and
functions, making it difficult to translate \safe policies into their
policy model.

Geode is a simple proxy for storage systems that do not manipulate
user data. It is inspired by previous systems that leverage a trusted
hardware platform to build secure storage out of untrusted
components~\cite{maheshwari00:_how_to_build_trust_datab,
  chen08:_overs, weinhold08:_vpfs, weinhold11:_jvpfs}.
More complex options that enable computation on stored data include
IFDB~\cite{schultz13:_ifdb} and CryptDB~\cite{popa11:_crypt}.

Magma builds on TaintDroid, a binary instrumentation tool for
programmers to find leaks of tainted data on Android applications.
TaintDroid has a different goal: it aims only to \emph{detect} flows,
not to stop them. As a result, it does not have a notion of policies
-- it tracks only a single bit of taint -- nor any enforcement
mechanism. TaintDroid is also a single-node system; its flow tracking
ends at the boundary of a single mobile device.

Our \safe framework is inspired by public key infrastructure (PKI).
While PKI has its issues (e.g., too many certificate authorities,
authorities issuing certificates that they do not own), there have
been efforts to address them (e.g., http public key
pinning~\cite{kranch15:_upgrad_https}). We hope that \safe will avoid
many of these pitfalls by relying on attestation instead of
authorities.  In particular, three factors differentiate \safe from
PKI: (1) it is more difficult for a system to become a trusted \safe
enforcement system than a certificate authority, (2) users can inspect
the code of open-source \safe enforcement systems and (3) mobile OSes
will not verify a cloud service as \safe unless it can attest that it
is running a trusted \safe enforcement system.

%% file: conclusion.tex

This paper introduced the \safe framework for distributed mobile apps.
\safe provides a system guarantee that user policies will be enforced
on sensitive user data no matter where it flows.  The framework relies
on \safe enforcement systems to provide this guarantee without
requiring application modifications.

The \safe framework leverages mobile OSes to vet cloud services.
Using attestation, a user's \safe mobile OS can verify that a cloud
service is running a trusted systems stack and a \safe enforcement
system.  Once verified, the mobile OS can trust the cloud service to
enforce \safe policies even if it is running untrusted applications.

This paper presents three \safe enforcement systems: Agate, a mobile
OS; Magma, a distributed runtime system; and Geode, a distributed
storage proxy.  Using these three systems, we were able to run several
unmodified applications across mobile devices and cloud server and
enforce \safe policies across the entire application.  Our results
demonstrate that the \safe framework is a practical way for users to
create a chain of trust from their mobile devices to the cloud.